\documentclass[12pt]{iopart}

\usepackage{iopams}  
\usepackage{graphicx}
\usepackage{xcolor}
\usepackage{ulem}
\usepackage{amssymb}

\newcommand{\bq}{\begin{equation}}
\newcommand{\eq}{\end{equation}}
\newcommand{\bqn}{\begin{eqnarray}}
\newcommand{\eqn}{\end{eqnarray}}
\newcommand{\nb}{\nonumber}
\newcommand{\lb}{\label}

\graphicspath{{Figuras/}}

\begin{document}
	
\title{Static Cylindrical Symmetric Solutions in the Einstein-Aether Theory}

\author{R. Chan $^{1}$ and M. F. A. da Silva $^{2}$} 

\address{$^{1}$Coordena\c{c}\~ao de Astronomia e Astrof\'{\i}sica, 
	Observat\'orio Nacional (ON), Rua General Jos\'e Cristino, 77, S\~ao Crist\'ov\~ao, CEP 20921-400, 	Rio de Janeiro, RJ, Brazil.\\
	$^{2}$Departamento de F\'{\i}sica Te\'orica,
	Instituto de F\'{\i}sica, Universidade do Estado do Rio de Janeiro (UERJ),
	Rua S\~ao Francisco Xavier 524, Maracan\~a,
	CEP 20550-900, Rio de Janeiro, RJ, Brazil.}
	
\ead{chan@on.br, mfasnic@gmail.com}

\begin{abstract}
In this work we present all the possible solutions for a static
cylindrical symmetric spacetime in the Einstein-Aether (EA) theory. 
As far as we know, this is the first work in the literature 
that considers cylindrically symmetric solutions in the theory of EA.
One of these solutions is the generalization in EA theory of the 
Levi-Civita (LC) spacetime in General Relativity (GR) theory.
We have shown 
that this generalized LC solution has unusual geodesic properties,
depending on the parameter $c_{14}$ of the aether field. The circular
geodesics are the same of the GR theory, no matter the values of $c_{14}$.
However, the radial and $z$ direction geodesics are allowed only for certain
values of $\sigma$ and $c_{14}$. The $z$ direction geodesics are restricted to an interval of $\sigma$ different from those predicted by the GR and the radial geodesics 
show that the motion  is confined between the origin and a maximum radius. The latter is not affected by the aether field but the velocity and acceleration of the test particles are Besides, for $0\leq\sigma<1/2$, when the cylindrical symmetry is preserved, 
this spacetime is singular at the axis $r=0$, although for $\sigma>1/2$ exists interval of $c_{14}$ where the spacetime is not singular, which
is completely different from that one obtained with the GR theory, 
where the axis $r=0$ is always singular.
\end{abstract}

\section{Introduction}

The landmark in the study of cylindrical symmetry in General Relativity (GR) was the pioneering work of Levi-Civita \cite{101}.
The difference between Newtonian and Einsteinian gravitational is striking
in this simple case that describes the vacuum field outside an infinite static cylinder of matter. In its general relativistic form, this solution contains two independent parameters \cite{102,103,104,106}, one describing the Newtonian energy per unit length of the source and the other related to the angular defect, in variation with its Newtonian counterpart that displays only the first parameter. The importance of the second emerges from its global topological meaning, since it cannot be removed by scale transformations \cite {102,118,119}. It produces a gravitational analogue for the Aharanov-Bohm effect that allows an unobservable quantity (Newtonian) (the constant potential beyond Newtonian potential) to become observable in relativistic theory through angular deficit chains \cite{118,119,105,132,149, Muriano, LewisWeyl}.
However, the first independent parameter, understood as Newtonian mass per unit length for small densities of matter, whose interpretation for higher mass densities is very difficult. In these densities, there are several apparently contradictory obstacles and properties, allowing for different possible interpretations (see a discussion in \cite {102,104,507}). 

Later, Linet \cite {160} and Tian \cite {170} presented LC space-time generalization to include the cosmological constant $ \Lambda $. It has been shown that the presence of the cosmological constant dramatically modifies space-time \cite{170,100,180,190,200,221,500}, such as, for example, its conformal properties.

In the context of GR theory, symmetrical cylindrical spacetimes are of great interest, as they allow the study of a wide range of physical systems, some of them exhibiting characteristics related to intrinsic symmetry (see, for example, \cite {100} and its references). The introduction of stationarity in a symmetrical cylindrical spacetime was performed by Lewis \cite {31}, obtaining two new  independent parameters. Furthermore, it has been shown that rotation gives rise to two families of space-time, one with a flat Minkowski space-time limit and the other without \cite {33,34,35,36}.
Later, Krasinski \cite {Krasinski} and Santos \cite {Santos} introduced the cosmological constant in this space-time.

All of these spacetime have been extensively studied for their geometric properties, their limits, their geodesics and particular sources. The results, some of them very strange, still need to be interpreted (such as the relationship between the metrics LC, Gamma and Schwarzschild \cite {Herrera1, Herrera2}), can be found in the articles cited here. 
\textcolor{black}{Besides, the
reference \cite{Bronnikov2020} can help the readers to see the present understanding of cylindrical systems in GR theory with cylindrical systems.}

The field equations of the time dependent vacuum with cylindrical symmetry were obtained by Einstein and Rosen \cite{Einstein}. They described the outer space-time for a collapsing cylindrical source \cite {229,217}, producing gravitational waves. In fact, cylindrical symmetry is the most simple symmetry capable to produces gravitational waves.

In another direction, in an attempt to overcome the difficulty imposed by the GR theory
in the quantization of gravitation, many researchers proposed several alternative theories. 
Among many of them, we will focus only in the EA theory, which considers 
the existence of a vector field that defines a preferred frame in the spacetime 
\cite{JacobsonMattingly2001}, \cite{Jacobson2009}. This theory was allowed to break the Lorentz's local symmetry but maintaining the general covariance and the second-order field equations. 
Since then, many studies have been developed to investigate the consequences of this theory, 
for example, in the field of cosmology \cite{CarrollLim2004}, \cite{CampistaEtAl}, \cite{ChanEtAlFried} and in the study of gravitational collapse \cite{ElingJacobson2006}, \cite{CroppLiberatiMohdVisser2014}, \cite{DingWangWang2015}, 
 \cite{BarausseSotiriouVega2016}, \cite{DingWangWangZhu2016}, \cite{ZhuWuJamilJusufi2019}, \cite{ChanEtAlSchw},
all of them restricted to the spherical symmetry. As cylindrical symmetry shows unusual behaviors in the context of the GR, we think that it is important to explore  it also in this new theory. 

Another motivation is related with the recent observations of gravitational waves in GR theory. Since then,
several works studied the existence of these waves in EA theory \cite{Abbott2016} \cite{Oost2018}\cite{Zhang2020}.
It is well known that the cylindrical symmetry is the simplest one capable of generating gravitational waves. 
Besides the gravitational waves, the cylindrical symmetry properties could better explain 
the observable acceleration of high energy jets in galaxies \cite{Gariel2017}). 
In another paper \cite{500}, one of us, considering a cylindrically symmetric source in the presence of positive cosmological constant, suggested to be  possible to  model a jet emerging outward, from the center of an galactic disc, and perpendicular to it. The confinement of the material, far away from the environment of the galaxy,  depends on the value of the cosmological constant. Other authors also identified some kind of confinement in cylindrically symmetric fluids \cite{Steadman}, \cite{Opher}.

Thus, it would be worth to search cylindrical symmetric solutions starting with the static vacuum solution, similar to the LC solution in GR. This solution, even in GR, has physical and geometric properties very different from those of Newtonian gravity, as already described above. 
In addition, it does not have event horizon, differently from 
spherically symmetric vacuum static vacuum.

Thus, in this pioneer work we find all the solutions for the static cylindrically symmetric vacuum in the EA theory and explore some of its properties. We show that there exist
a generalization of the LC spacetime \cite{101}, but the presence of the unitary vector field can change some geometric aspects, since it appears explicitly in the metric.

The paper is organized as follows.  In Section 2 we present the field equations in EA. 
In Section 3 we show the general solutions of these equations, and the generalization
of the LC spacetime. In Section 4 we present the geodesic properties of this generalization
of the LC spacetime. Finally in Section 5 in discuss our results.

\section{Field Equations in the EA theory }

The general action of the EA theory, in a background where the  metric signature is $({-}{+}{+}{+})$ and the units are chosen so that the speed of light defined by the metric $g_{ab}$ is unity, is given by 
\bq 
S =  \frac{1}{16\pi G}\int \sqrt{-g}~(R+L_{\rm aether}+L_{\rm matter}) d^{4}x,
\label{action}
\eq
where, the first term defined by $R$ is the usual Ricci scalar, and $G$ the EA coupling constant.
The second term, the aether Lagrangian is given by
\bq 
L_{\rm aether} =  \frac{1}{16\pi G} [-K^{ab}{}_{mn} \nabla_a u^m
\nabla_b u^n +
\lambda(g_{ab}u^a u^b + 1)],
\lb{LEAG}
\eq
where the tensor ${K^{ab}}_{mn}$ is defined as
\bq {{K^{ab}}_{mn}} = c_1 g^{ab}g_{mn}+c_2\delta^{a}_{m} \delta^{b}_{n}
+c_3\delta^{a}_{n}\delta^{b}_{m}-c_4u^a u^b g_{mn},
\lb{Kab}
\eq
being the $c_i$ dimensionless coupling constants, and $\lambda$
a Lagrange multiplier enforcing the unit timelike constraint on the aether, and 
\bq
\delta^a_m \delta^b_n =g^{a\alpha}g_{\alpha m} g^{b\beta}g_{\beta n}.
\eq

Finally, the last term, $L_{\rm matter}$ is the matter Lagrangian, which depends on the metric tensor and the matter field.

In the weak-field, slow-motion limit EA theory reduces to Newtonian gravity with a value of  Newton's constant $G_{\rm N}$ related to the parameter $G$ in the action (\ref{action})  by  { \cite{Garfinkle2007}},
\bq
G = G_N\left(1-\frac{c_1+c_4}{2}\right).
\lb{Ge}
\eq
 
Note that if $c_1=-c_4$ the EA coupling constant $G$ becomes the Newtonian coupling constant $G_N$, without {necessarily} imposing $c_1=c_4=0$.

The field equations are obtained by extremizing the action with respect to independent  variables of the system. The variation with respect to the Lagrange multiplier $\lambda$ imposes the condition that $u^a$ is a unit timelike vector, thus 
\bq
g_{ab}u^a u^b = -1,
\lb{LagMul}
\eq
while the variation of the action with respect $u^a$, leads to \cite{Garfinkle2007}
\bq
 \nabla_a J^a_b + c_4 a_a \nabla_b u^a + \lambda u_b = 0,
\eq
where,
\bq
J^a_m=K^{ab}_{mn} \nabla_b u^n,
\eq
and
\bq
a_a=u^b \nabla_b u_a.
\lb{aa}
\eq
The variation of the action with respect to the metric $g_{mn}$ gives the dynamical equations,
\bq
G^{Einstein}_{ab} = T^{aether}_{ab} +8 \pi G  T^{matter}_{ab},
\lb{EA}
\eq
where 
\bqn
G^{Einstein}_{ab} &=& R_{ab} - \frac{1}{2} g_{ab} R, \nb \\
T^{aether}_{ab}&=& \nabla_c [ J^c\;_{(a} u_{b)} + u^c J_{(ab)} - J_{(a} \;^c u_{b)}] - \frac{1}{2} g_{ab} J^c_d \nabla_c u^d+ \lambda u_a u_b  \nb \\
& & + c_1 [\nabla_a u_c \nabla_b u^c - \nabla^c u_a \nabla_c u_b] + c_4 a_a a_b, \nb \\
T^{matter}_{ab} &=&  \frac{- 2}{\sqrt{-g}} \frac{\delta \left( \sqrt{-g} L_{matter} \right)}{\delta g_{ab}}.
\lb{fieldeqs}
\eqn

Later, when we solve the field equations (\ref{EA}), we do take into consideration the equations 
(\ref{LagMul})-(\ref{aa}) in the process of simplification. Thus in the paper (as in the equations (\ref{G11})-(\ref{G44}) below) we seem to solve only the dynamical equations, but in fact we are also solving the equations arising from the variations of the action with respect $\lambda$ and $u^a$.

In a more general situation, the Lagrangian of GR theory is recovered, if and only if, the coupling constants are identically null, e.g., $c_1=c_2=c_3=c_4=0$, 
{considering} the equations 
{(\ref{Kab}) and (\ref{LagMul})}.

We will assume 
the most general static cylindrical symmetric is given by
\bq
ds^2= -A(r) dt^2 +B(r) \left(dr^2 + dz^2\right) +C(r) d\phi^2 ,
\lb{ds2}
\eq
and we assume the indices of the Riemann e Einstein tensors as $(1,2,3,4)$ corresponding to the coordinates $(r,\phi,z,t)$, respectively.

The components of the Riemann tensor are given by
\bq
R_{1212} = -\frac{1}{4B C} (2 C'' B C-B' C' C-C'^2 B),
\eq
\bq
R_{1313} = -\frac{1}{2B} (B'' B-B'^2),
\eq
\bq
R_{1414} = \frac{1}{4B A} (2 A'' B A-A' B' A-A'^2 B),
\eq
\bq
R_{2323} = -\frac{ B' C'}{4B},
\eq
\bq
R_{2424} = \frac{ A' C'}{4B},
\eq
\bq
R_{3434} = \frac{ A' B'}{4B}.
\eq

An important quantity to be computed is the Kretschmann scalar. For this metric, it is  given by
\bqn
K=\frac{1}{4B^6 A^4 C^4}&& \left[ 4 A^4 C^4 B''^2 B^2-8 A^4 C^4 B'' B B'^2+4 A^4 C^4 B'^4+ \right.\nb\\
&&\left. 2 A'^2 B'^2 B^2 A^2 C^4+2 B'^2 C'^2 B^2 A^4 C^2+4 B^4 C^4 A''^2 A^2- \right.\nb\\
&&\left. 4 B^3 C^4 A'' A^2 A' B'-4 B^4 C^4 A'' A A'^2+2 B^3 C^4 A'^3 B' A+ \right.\nb\\
&&\left. B^4 C^4 A'^4+A'^2 C'^2 B^4 A^2 C^2+4 B^4 A^4 C''^2 C^2-\right.\nb\\
&&\left. 4 B^3 A^4 C'' C^2 B' C'-4 B^4 A^4 C'' C C'^2+2 B^3 A^4 B' C'^3 C+\right.\nb\\
&&\left. B^4 A^4 C'^4\right].
\lb{K}
\eqn

In accordance with equation (\ref{LagMul}), the aether field is assumed unitary, timelike and constant, chosen as
\bq
u^a=(A^{-\frac{1}{2}},0,0,0).
\eq

Assuming (\ref{ds2}), we compute the different terms in the field equations (\ref{fieldeqs}), giving 
\bqn
G^{aether}_{11} = \frac{1}{8B C A^2}&& \left(2 B' C' A^2+2 A' B' A C+2 A B C' A'+\right.\nb\\
&&\left.A'^2 B C c_{14}\right)=0,
\lb{G11}
\eqn
\bqn
G^{aether}_{22} = -\frac{C}{8B^3 A^2}&& \left(4 B'^2 A^2-4 B'' B A^2+2 A'^2 B^2-4 A'' B^2 A+\right.\nb\\
&&\left.A'^2 B^2 c_{14}\right) =0,
\lb{G22}
\eqn
\bqn
G^{aether}_{33} = &&-\frac{1}{8B C^2 A^2}\left(2 B' C' C A^2+2 B' A' C^2 A+2 C'^2 B A^2-4 C'' C B A^2+\right.\nb\\
&&\left.2 A'^2 C^2 B-4 A'' C^2 B A-2 A B C' A' C+A'^2 C^2 B c_{14}\right) =0,
\lb{G33}
\eqn
\bqn
G^{aether}_{44} = &&\frac{1}{8B^3 C^2 A}(2 C'^2 B^2 A^2-4 C'' C B^2 A^2+4 B'^2 C^2 A^2-4 B'' C^2 B A^2+\nb\\
&&3 B^2 C^2 A'^2 c_{14}-4 B^2 C^2 A'' A c_{14}-2 B^2 C A' C' A c_{14})=0,
\lb{G44}
\eqn
where the symbol prime denotes the derivative in relation to the coordinate $r$ and $G^{aether}_{\mu\nu}=G^{Einstein}_{\mu\nu}-T^{aether}_{\mu\nu}$.
We assume also $c_{14}=c_1+c_4$.

\section{{\bf The General Solutions}}

Solving simultaneously the field equations (\ref{G11})-(\ref{G44}) we obtain three possible
solutions.

\subsection{{\bf Case 1:}}

\bq
A = C_4,
\eq
\bq
B=e^{C_{1} r} C_{2},
\eq
\bq
C= C_{3},
\eq
\bq
R_{\alpha\beta\delta\gamma}=0,
\eq
\bq
K=0.
\eq
Hereinafter, $C_1$, $C_2$, $C_3$, $C_4$ and $C_5$ are arbitrary constants of integration. In this case, since Riemann tensor vanishes, we can
conclude that we have a Minkowski spacetime, but not globally. 
This can be clearly seen with the coordinates transformation:
\bqn
t & = & \frac{1}{\sqrt{C_{4}}}\bar{t}, \nb\\
r & = & \frac{2}{C_{1}}\ln{\frac{C_{1}\bar{r}}{2\sqrt{C_{2}}}} .\nb 
\eqn

In terms of these new coordinates the metric (\ref{ds2}) becomes
\bq
ds^2= -d{\bar{t}}^2 +d{\bar{r}}^2 + \frac{{C_{1}}^2}{4C_{2}}{\bar{r}}^2 dz^2 +C_{3}d\phi^2 .
\lb{ds2_1}
\eq

Then we can say that  we have a metric of a static cosmic string  
as long as we interchange the role of the coordinates $z$ with $\phi$, that is, $z\rightarrow \bar{\phi}$ and $\phi\rightarrow \bar{z}$ and rescale the new coordinate $\bar{z}$ by the  absorption of the constant $C_{3}$. The range of the new angular coordinate $\bar{\phi}$ is now $0\leq\bar{\phi}\leq C_{1}\pi/\sqrt{C_{2}}$, showing an angular deficit of
$\delta=\pi(2-C_1/\sqrt{C_{2}})$.
The final metric is
 
\bq
ds^2= -d{\bar{t}}^2 +d{\bar{r}}^2 + d{\bar z}^2+\frac{{C_{1}}^2}{4C_{2}}{\bar{r}}^2 d{\bar{\phi}}^2 .
\lb{ds2_1a}
\eq

\subsection{{\bf Case 2:}}

\bq
A = C_4,
\eq
\bq
B=C_{1},
\eq
\bq
C= \frac{1}{4} C_{2}^2 r^2+\frac{1}{2} C_{2} C_{3}r+\frac{1}{4} C_{3}^2  = \left(\frac{C_{2}r}{2}+\frac{C_{3}}{2} \right)^2,
\eq
\bq
R_{\alpha\beta\delta\gamma}=0,
\eq
\bq
K=0.
\eq
Also in this case, since Riemann tensor vanishes, we can
conclude that we have  again a locally Minkowski spacetime. 
The metric can be put in the form
\bq
ds^2= -d{\bar{t}}^2 +d{\bar{r}}^2 + d{\bar z}^2+\frac{{C_{2}^2}}{4C_{1}}{\bar{r}}^2 d{\phi}^2 ,
\lb{ds2_2}
\eq
where we consider the coordinate transformations
\bqn
t & = & \frac{1}{\sqrt{C_{4}}}\bar{t}, \nb\\
r & = & \frac{1}{\sqrt{C_{1}}} \bar{r}-\frac{C_3}{C_2}, \nb\\ 
z & = & \frac{1}{\sqrt{C_{1}}}\bar{z},
\eqn
and the range of the coordinate $\phi$ is now 
$0\leq\phi\leq \pi\frac{C_{2}}{\sqrt{C_{1}}}$.

\subsection{{\bf Case 3:}}

\bq
A =  C_5(r+C_4)^{C_3},
\eq
\bq
B=C_1 (r+C_4)^{-\frac{1}{4} (4-2 C_3+C_3 c_{14}) C_3}
\eq
\bq
C=\frac{C_2}{C_5 C_3^2} (r+C_4)^{-C_3+2},
\eq
where $C_1$, $C_2$, $C_3$, $C_4$ and $C_5$ are arbitrary constants of integration.
Choosing $C_1 = 1$, $C_2=16\sigma^2 a^{-2}$, $C_3 = 4 \sigma$, $C_4 = 0$, $C_5 = 1$, 
where $a$ is an arbitrary constant, we can write the metric as a generalization of 
the LC metric, i.e.
\bq
A = r^{4 \sigma},
\eq
\bq
B=r^{4\sigma(2\sigma-1-\sigma c_{14})},
\eq
\bq
C= a^{-2} r^{2-4 \sigma},
\eq
\bq
ds^2= -r^{4 \sigma} dt^2 +r^{4\sigma(2\sigma-1-\sigma c_{14})} \left(dr^2 + dz^2\right) +a^{-2} r^{2-4 \sigma} d\phi^2,
\lb{ds3}
\eq
Note that it reduces to the LC metric if $c_{14}=0$, where $\sigma$ is associated with the linear energy density of the cylindrical source of the vacuum spacetime and $a$ is connected with an angular deficit.
The Kretschmann scalar is given by
\bqn
K=&&16 r^f\sigma^2\times\nb\\ &&[(16\sigma^4-8\sigma^3+3\sigma^2) c_{14}^2+ (64\sigma^3-64\sigma^4+4\sigma-24\sigma^2) c_{14}-\nb\\ &&96\sigma^3-24\sigma+64\sigma^4+64\sigma^2+4],
\eqn
\bq
r_{sing} = 0,\; \quad{\rm \quad for\quad f<0},
\eq
where
\bq
f={8\sigma-16\sigma^2+8\sigma^2 c_{14}-4}.
\eq

\begin{figure}[!ht]
\includegraphics[width=\linewidth]{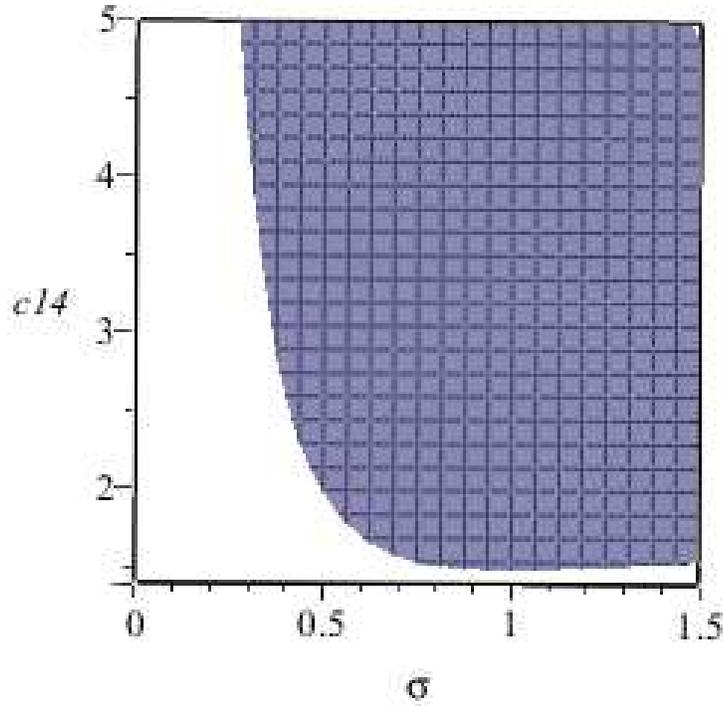}
\caption{This figure shows the different combinations between $\sigma$ and $c_{14}$ determining the sign of the power $f$. The gray region furnishes $f\ge 0$ and white region furnishes $f<0$.}
\label{f4}
\end{figure}

Since the exponent of $r$, that is, $f$, is always negative for
$\sigma<1/2$ and for $c_{14}<2$, then we have $\lim_{r\rightarrow r_{sing}}K=+\infty$ and the spacetime is singular at $r=0$.
On the other hand, for $\sigma >1/2$, $f$ can be positive for some combinations of $c_{14}$ and $\sigma$. The gray region in Figure \ref{f4} includes the combinations of $c_{14}$ and $\sigma$ for what $f$ is positive and the white region, the combinations of $c_{14}$ and $\sigma$ for what $f$ is negative. Then, depending on the choice of $\sigma$ and $c_{14}$ in the range $\sigma>1/2$ and $c_{14}>3/2$, or $\sigma<1/2$ and $c_{14}>2$, the spacetime can be not singular at $r=0$. This result is completely different from that one obtained with the GR theory, where the axis $r=0$ is always singular for all values of $\sigma$ different from $0$, $1/2$ and $\infty$, since the exponent $f$ is always negative if $c_{14}=0$ \cite{HerreraSantosTeixeira}. Then, we can say that the aether interferes in the global structure of the spacetime. In order to make sure of this result, we also list, in the Appendix A, the 18 scalar polynomial invariants of the Riemann tensor furnished by the GRTensorII package using Maple 16 \cite{Carminati1991}, \cite{Zakhary1997}. We can notice that the power of $r$ in all of the non null invariants is always proportional to the function $f$. This means that the same analysis made for the Kretschmann scalar is valid for all these invariants.

In fact, in order to assure the cylindrical symmetry in the generalized LC metric (\ref{ds3}), the parameter $\sigma$ must be restrict to range $0\leq\sigma\leq 1/2$. For $1/2<\sigma<\infty$, the azimuthal variable $z$ and the angular variable $\phi$ must interchange their roles  \cite{HerreraSantosTeixeira}.

It is interesting to evaluate also the limits of the solutions when $\sigma=0$ and $\sigma=1/2$. For the first one, the solution corresponds to the Minkowski solution, although only locally because the presence of the angular deficit change some of the global properties of the metric. On the other hand, for $\sigma=1/2$, the aether field vector prevents a plane symmetric solution, differently from the analogue in the GR theory, due to the presence of the parameter $c_{14}$, as can be seen in the resulting metric
\bq
ds^2= -r^2 dt^2 +r^{-c_{14}} \left(dr^2 + dz^2\right) +a^{-2} d\phi^2.
\lb{ds4}
\eq
Thus, the symmetry of the spacetime can be modified by the vector field since we do not have anymore the Rindler flat spacetime, whose sections for each time "$t$" have planar symmetry, as foreseen by the GR theory  \cite{HerreraSantosTeixeira}.

\section{Geodesic equations}

Even in the GR theory there is a gap of studies about cylindrically symmetric sources, making it difficult to interpret the parameters of the vacuum solution. Then, many times, the researches make use of the geodetic movement of a test particle in order to study the effects generated by these in the external vacuum space. Considering the metric (\ref{ds3}), the geodesic equations are
\bqn
\ddot{t}+ \frac{4\sigma\dot{t}\dot{ r }}{ r } = 0,
\label{geot}
\eqn
\bqn
\ddot{ r } + 2\alpha\left(\frac{\dot{ r }^2-\dot{z}^2}{ r }\right)+\frac{1}{a^2}(2\sigma-1) r ^{(1-4\alpha-4\sigma)}\dot{\phi}^2+2\sigma r ^{(4\sigma-4\alpha-1)}\dot{t}^2 = 0,
\label{georho}
\eqn
\bqn
\ddot{z} +\frac{4\alpha\dot{z}\dot{ r } }{ r }= 0,
\label{geoz}
\eqn
\bq
\ddot{\phi} + \frac{2(1-2\sigma)\dot{ r }\dot{\phi}}{ r } = 0,
\label{geophi}
\eq
where $\alpha=\sigma[\sigma(2-c_{14})-1]$ and the dot denotes, hereinafter, differentiation with respect to the proper time $\tau$. 

In addition, from (\ref{geoz}), (\ref{geophi}), and considering (\ref{geot}), we can see that accelerated motions on $z$ and $\phi$ directions, in general, are allowed only if $\dot r \neq0$. In particular, if $\sigma=1/2$, for movements in the $\phi$ direction, or if $\sigma=1/(2-c_{14})$, for movements in the direction $z$, test particles are never accelerated. Comparing these results with those of the LC solution in the GR theory, we see here a clear difference between the two theories: if $c_{14}\neq 0$ movements in $z$ are affected by the vector field of the aether through $c_{14}$. Equation (\ref{georho}) reveals that this same dependence also occurs in the $ r $ direction.
On the other hand, there is none influence of the aether field in the $\phi$ direction.

Assuming $\dot  r  \ne 0$, equations (\ref{geot}), (\ref{geoz}) and (\ref{geophi}) can be solved furnishing

	\begin{equation}
	\label{tponto}
	\dot t=E  r ^{-4\sigma},
	\end{equation}
	\begin{equation}
	\label{zponto}
	\dot z=P_z  r ^{4\sigma[1-\sigma(2-c_{14})]},
	\end{equation}
	\begin{equation}
	\label{phiponto}
	\dot \phi=a^2L_z  r ^{2(2\sigma-1)},
	\end{equation}
where $E$, $P_z$ and $L_z$, are the total energy of the
test particle, its azimuthal linear momentum and its angular momentum about the $z$ axis,
respectively \cite{500}\cite{Brito2014}. Besides, following the method of the references 
\cite{500}\cite{Brito2014} we can also write
	\begin{equation}
	\label{rponto}
	\dot r^2=r ^{4\sigma^2(c_{14}-2)} \left\{ E^2 -r^{4 \sigma} \left[ 1 - P_z  r ^{4\sigma[1-\sigma(2-c_{14})]} - a^2L_z  r ^{2(2\sigma-1)} \right] \right\}
	\end{equation}

Next, we will detail the movement of test particles in each of the main directions.
Since the equations (\ref{tponto})-(\ref{phiponto}) assume that  $\dot  r  \ne 0$,
we have to return to the general geodesic equations (\ref{geot})-(\ref{geophi}),
in order to analyze the particular cases: circular, z-direction and radial geodesics.

\subsection{Circular geodesic}

For the circular geodesics we have the following conditions
\bq
r  =  r _0=constant,\; \dot  r  = 0,\; \ddot  r  = 0,\; \dot z = 0,\; \ddot z = 0.
\eq
Then, equations (\ref{geot}), (\ref{geoz}) and (\ref{geophi}) give us $\ddot t =0$, $\ddot z =0$ and $\ddot \phi =0$, while equation (\ref{georho}) furnishes
\bq
(2\sigma-1){\dot \phi}^2 =-a^2 \sigma { r _0}^{2(4\sigma-1)},
\eq
with
\bq
t(\tau)=n_2\;\tau+{n_3},
\eq
and
\bq
\phi \left( \tau \right) =\dot \phi\; \tau +{n_1},
\eq
where $n_1$, $n_2$ and $n_3$ are arbitrary integration constants. Thus, the squared angular 
velocity of a test particle is given by
\bq
\label{CircGeoLC}
\omega^2={\dot \phi}^2= \frac{2a^2 \sigma}{1-2\sigma}{ r _0}^{2(4\sigma-1)},
\eq
and the tangential velocity $W^\phi$ is given by $W^\phi= \omega/\sqrt{A}$, i.e., 
\begin{equation}
W^2=\frac{\omega^2}{A}=\frac{2\sigma}{1-2\sigma}.
\label{CircGeoLC1}
\end{equation}

We can see that it is exactly the same angular velocity for test particles in the LC spacetime in GR theory \cite{HerreraSantosTeixeira}, being admissible only in the range $0\leq\sigma<1/2$.  When $\sigma=0$ then $\dot \phi= 0$ or $W=0$, and when $\sigma=1/4$ then 
$\dot \phi= a$, remembering that the constant $a$ gives us a measure of the angular defect produced in outer time space by the cylindrical source 
\cite{104}, \cite{Muriano}, \cite{WdaSS}, or $W=1$, implying in a luminal velocity as already pointed out in \cite{HerreraSantosTeixeira}.

\subsection{z-direction geodesic}

Our interest now is to better investigate the conditions so that test particles can move exclusively in the z direction, that is,
parallel to the axis of symmetry, if this is possible. In this case, we have
\bq
 r  =  r _0=constant,\; \dot  r  = 0,\; \ddot  r  = 0,\; \dot \phi = 0,\; \ddot \phi = 0.
\eq
Then, equation (\ref{geophi}) is identically satisfied, while equations (\ref{geot}), (\ref{georho}) and (\ref{geoz})  furnish

\bqn
\dot z&=&\pm {\frac {\sqrt {- \left( 1-2\,\sigma+{c_{14}}\,\sigma \right) { r _{{0}}}^{4\,\sigma\, \left( 2-2\,\sigma+{c_{14}}\,\sigma \right) }}}{1-2\,\sigma+{c_{14}}\,\sigma}},\nb\\
&&=\pm \frac {\sqrt {\alpha \sigma { r _{{0}}}^{4(\sigma-\alpha)}}}{\alpha},
\lb{vz2}
\eqn

\bq
t(\tau)=n_2\;\tau+{n_3},
\eq
\bq
z(\tau)={\dot z}\,t(\tau)+{n_4},
\eq
where $\pm$ indicates movement toward positive $z$ coordinate ($+$) and toward negative $z$ coordinate ($-$). We have to assume $n_4=0$ in order to satisfy the field equations.

\begin{figure}[!ht]
	\includegraphics[width=\linewidth]{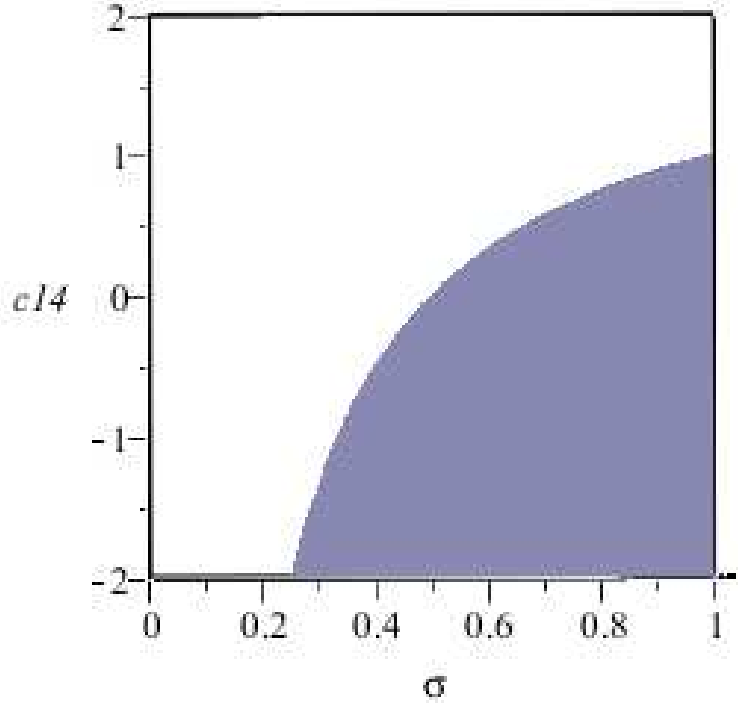}
	\caption{The figure shows a gray region indicating the combinations between $c_{14}$ and $\sigma$ for which $\alpha$ is positive.} 
	\label{alpha}
\end{figure}

We can note from equation (\ref{vz2}) that the $z$ velocity of the test particle
depends on $\alpha$. Figure \ref{alpha} shows, in the gray region, the combinations between $c_{14}$ and $\sigma$ for which $\alpha$ is positive, while the white region indicates the combinations for which $\alpha$ is negative and therefore $\dot z$ is imaginary, independent of the values of $ r $.

Thus, assuming the positive $z$ movement, we have two different possibilities:
\begin{enumerate}
\item {If $c_{14}>\frac{2\sigma-1}{\sigma}$ then $\alpha<0$. Thus, $\dot z$ is imaginary,}
\item {If $c_{14}<\frac{2\sigma-1}{\sigma}$ then $\alpha>0$. Thus, $\dot z$ decreases with $ r $}.
\end{enumerate}

Here we have an apparent  difference between the EA and GR theory, since in GR theory there are no trajectories parallel to the axis of symmetry for $0\leq\sigma\leq 1/2$, as seen in equation (\ref{vz2}) if $c_{14}=0$ \cite{CelerierEtAl}. In fact, in order to assure the cylindrical symmetry in the LC metric, in GR or EA theory, the parameter $\sigma$ must be restrict to range $0\leq\sigma\leq 1/2$, as mentioned before \cite{HerreraSantosTeixeira}. Here we show
that the presence of the aether field, represented by the parameter $c_{14}$, modifies this result, allowing a trajectory like this, for a range of values of $\sigma$ within the  interval indicated by the gray region in Figure \ref{alpha}, but only if $c_{14}$ can be negative.  This unexpected result, motions parallel to the symmetry axis for $0\leq\sigma\leq 1/2$, shows that combinations of the parameters $ c_1 $ and $ c_4 $ such that lead to $ c_ {14} <0 $ are undesirable, which is in accordance with what we know about the EA theory in spherical symmetry. Many efforts have been made in order to establish limits on the parameters $c_1$, $c_2$, $c_3$ and $c_4$ to ensure that the theory reproduces the observational results known at the weak field limit. In this way, Foster and Jacobson \cite{FosterJacobson2006}, using the PPN (Parameterized Post-Newtonian) formalism established narrow limits for these parameters. Regarding the parameters $c_1$ and $c_4$, present in the solutions studied here by us, through the combination $c_{14}=c_1+c_4$, they concluded that the condition $c_{14}>0$ is necessary to guarantee the absence of ghosts in the perturbed EA action. Therefore, we conclude that the $z$-direction geodesics does not exist in a healthy EA theory, coinciding with that one in GR theory.

\subsection{Radial geodesic}

We can ask what happens with a test particle on a purely radial path. In this case we must have
\bq
\dot z = 0,\; \ddot z = 0,\; \dot \phi = 0,\; \ddot \phi = 0.
\eq
Then, equations (\ref{geoz}) and (\ref{geophi}) are identically satisfied, while equations (\ref{geot}) and (\ref{georho}) becomes
\bq
\ddot r +2\alpha\frac{{\dot r }^2}{ r }+2\sigma r ^{4\sigma^2(c_{14}-2)-1}=0,
\lb{ddotr}
\eq
\bq
t(\tau)={n_5}+{n_6}\int \!  r \left( \tau \right)  ^{-4\,\sigma}{d\tau},
\eq
and as an example see Figure 4.

This can be integrated through a variable transformation in order to lower the order of the integral, so that the solution obtained is
\bq
\dot r=v_r=\sqrt{r^{4\,{\sigma}^{2} \left( {c_{14}}-2 \right) } \left( 
n_6^{2}+n_7\;{r}^{4\,\sigma} \right)},
\lb{vr}
\eq

where $n_7$ is another arbitrary integration constant.
We can also rewrite the equation (\ref{rponto}) as
\begin{equation}
	\label{rpontoa}
	\dot r^2=\frac{1}{AB} \left[ E^2 - V(r) \right],
	\lb{dotr}
	\end{equation}
where $V(r)$ is the potential
\bqn
V(r)&=&A - P_z^2\frac{A}{B} - L_z^2\frac{A}{C}\nb\\
&=&r^{4 \sigma} \left[ 1 -  P_z^2 r ^{4\sigma[1-\sigma(2-c_{14})]} - L_z^2  r ^{2(2\sigma-1)} \right].
\lb{Vr}
\eqn
Using the equation (\ref{rponto}) and assuming in this case $P_z=0$ and $L_z=0$ we can identify in the equation (\ref{vr}) the constant $n_6$ as the total energy $E$, $n_7=-1$ and the potential as $V(r)={r}^{4\,\sigma}$ (see Figure 5).

Here we can see that if we choose  $\sigma=0$ in the equation (\ref{georho}), 
as already showed, we fall into  trivial Minkowski's spacetime case and, in this case,
$\dot r =\sqrt{{E}^2-1}$
represents the initial radial velocity of the particle that will remain constant. Therefore, we must assure that 
$1<{E}<\sqrt{2}$, in order to assures real velocities lower than the vacuum light velocity.

Note that the radial velocity, given by (\ref{vr}), is null at the radius
\bq
r_{max} = E^{\frac{1}{2\sigma}}.
\eq
The test particles will be restricted to a region of space, between the axis of symmetry and $r_{max}$. Here we can see that this maximum radius does not depend on the parameter $c_{14}$, although the velocity and the acceleration in the radial direction of the test particle is modified by the aether vector field, as can be seen in the Figure 3. 

\begin{figure}[!ht]
\centering
	\includegraphics[width=7cm]{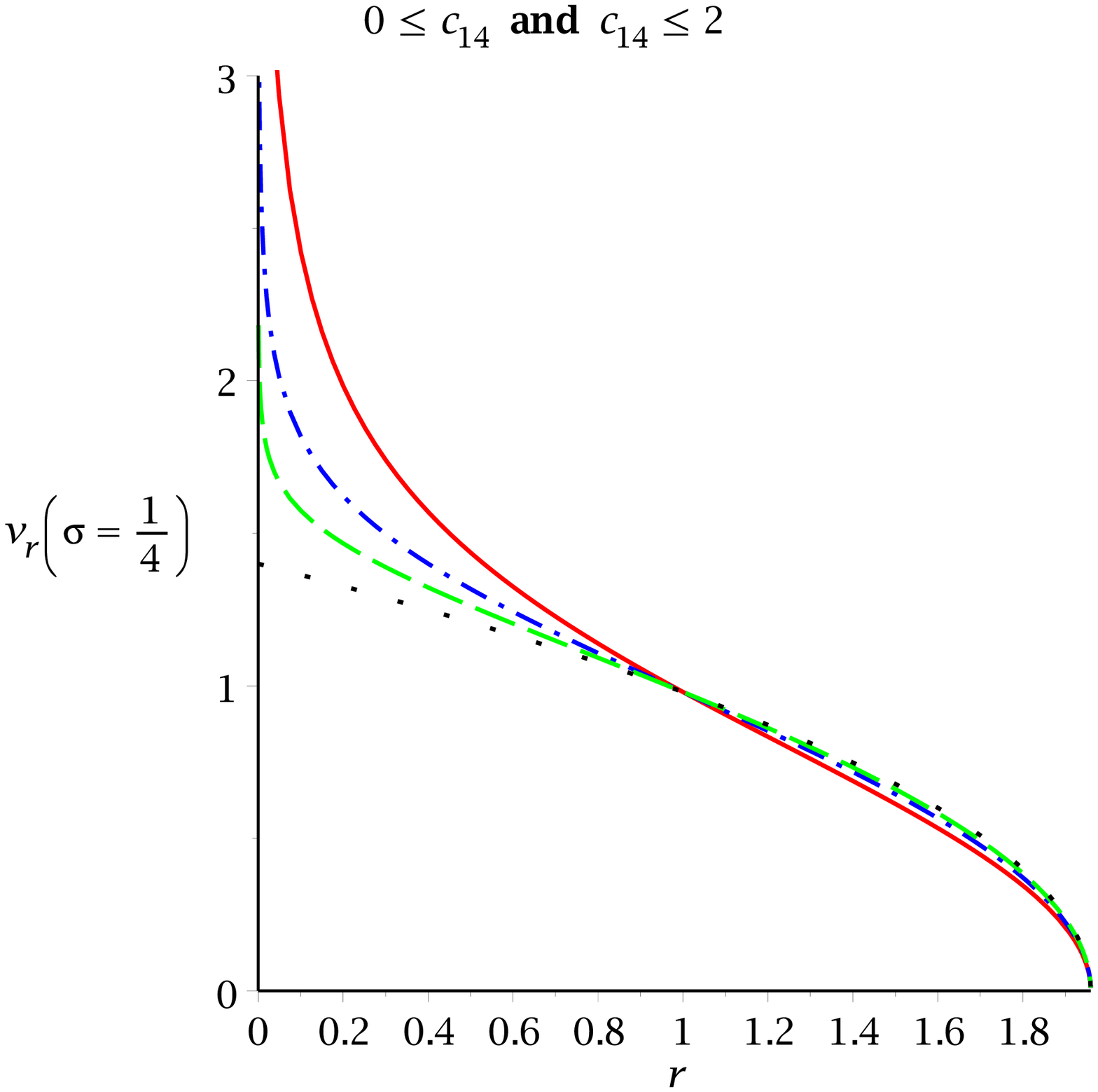}
	\includegraphics[width=7cm]{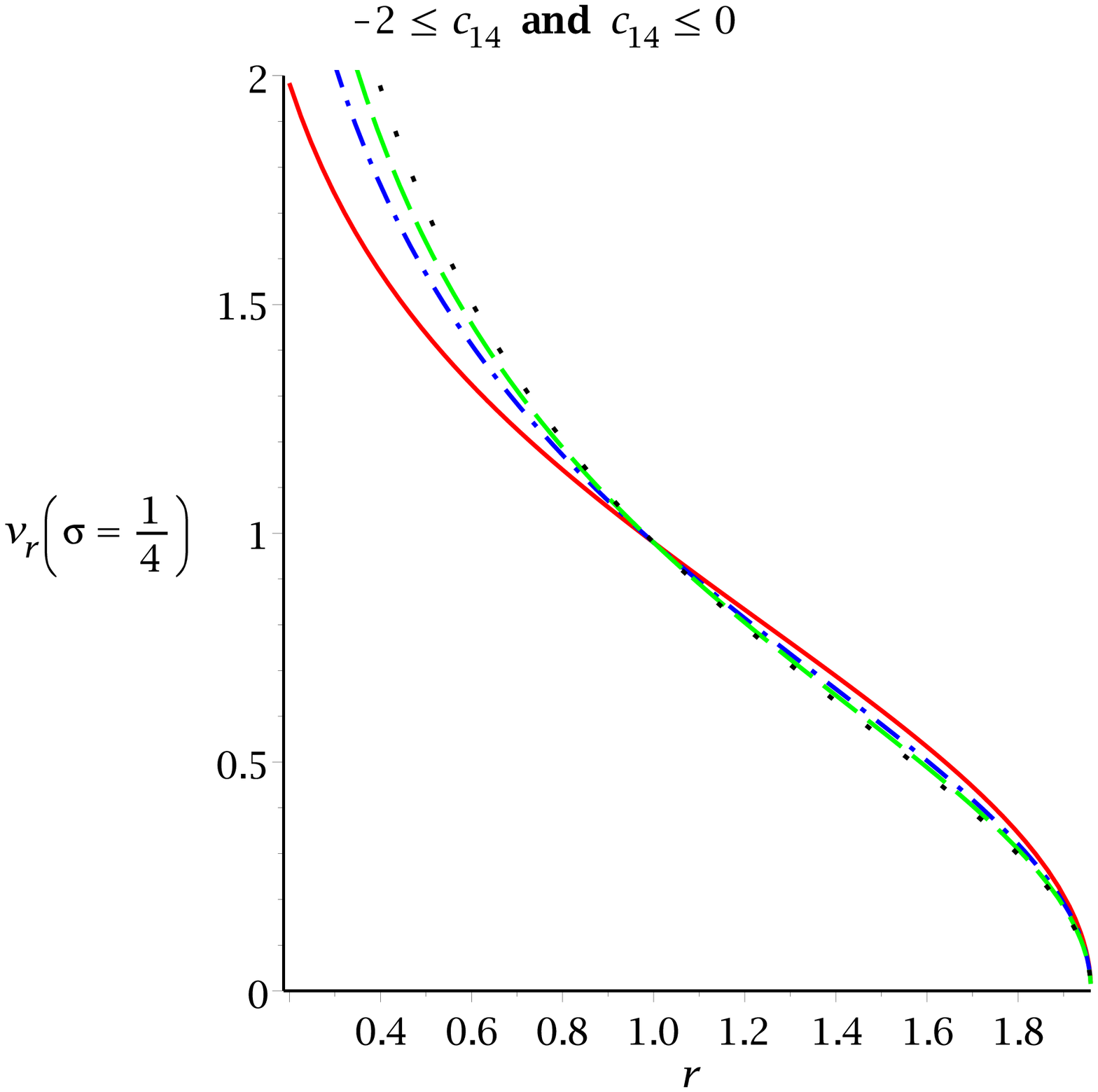}
	\includegraphics[width=7cm]{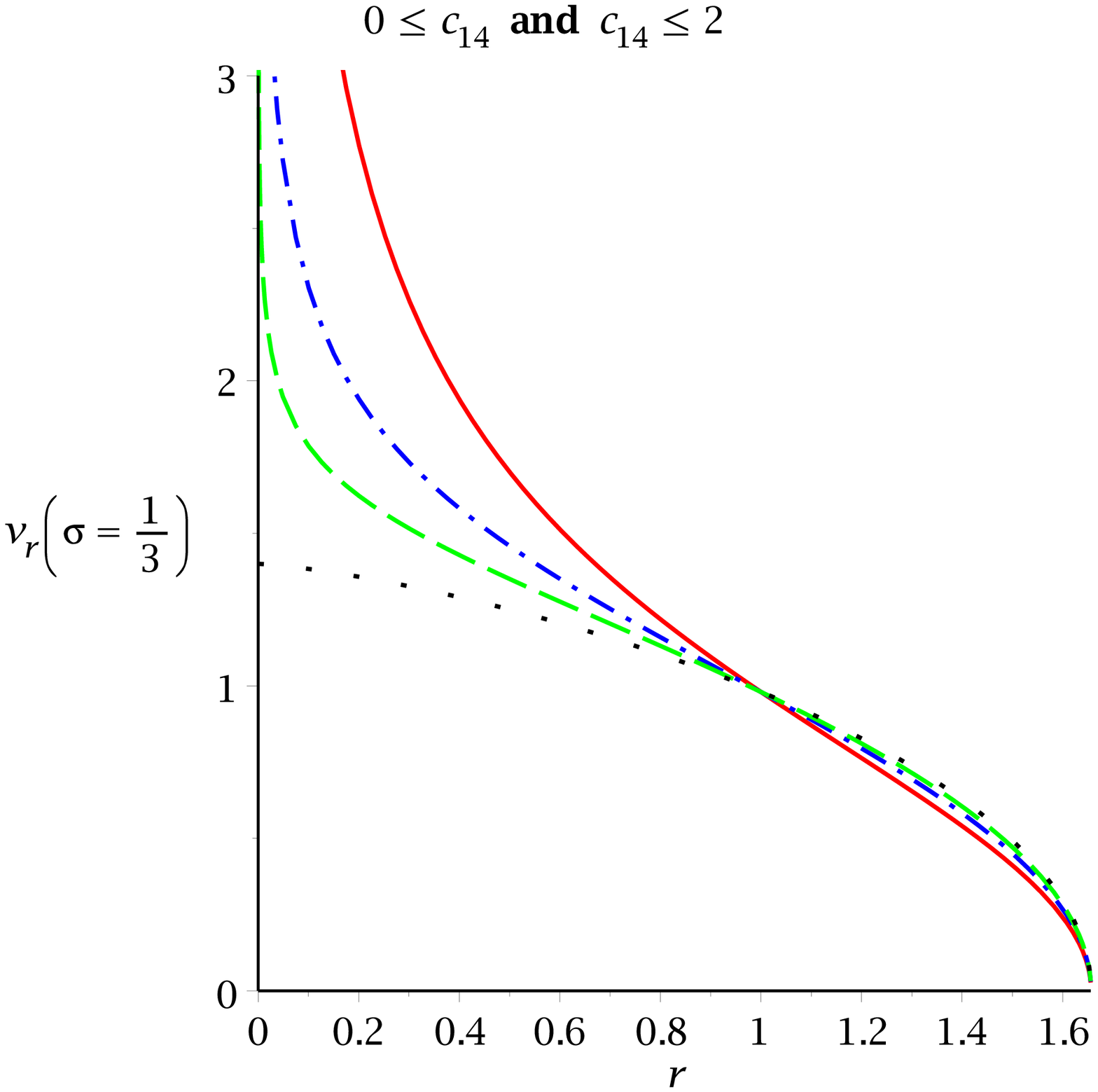}
	\includegraphics[width=7cm]{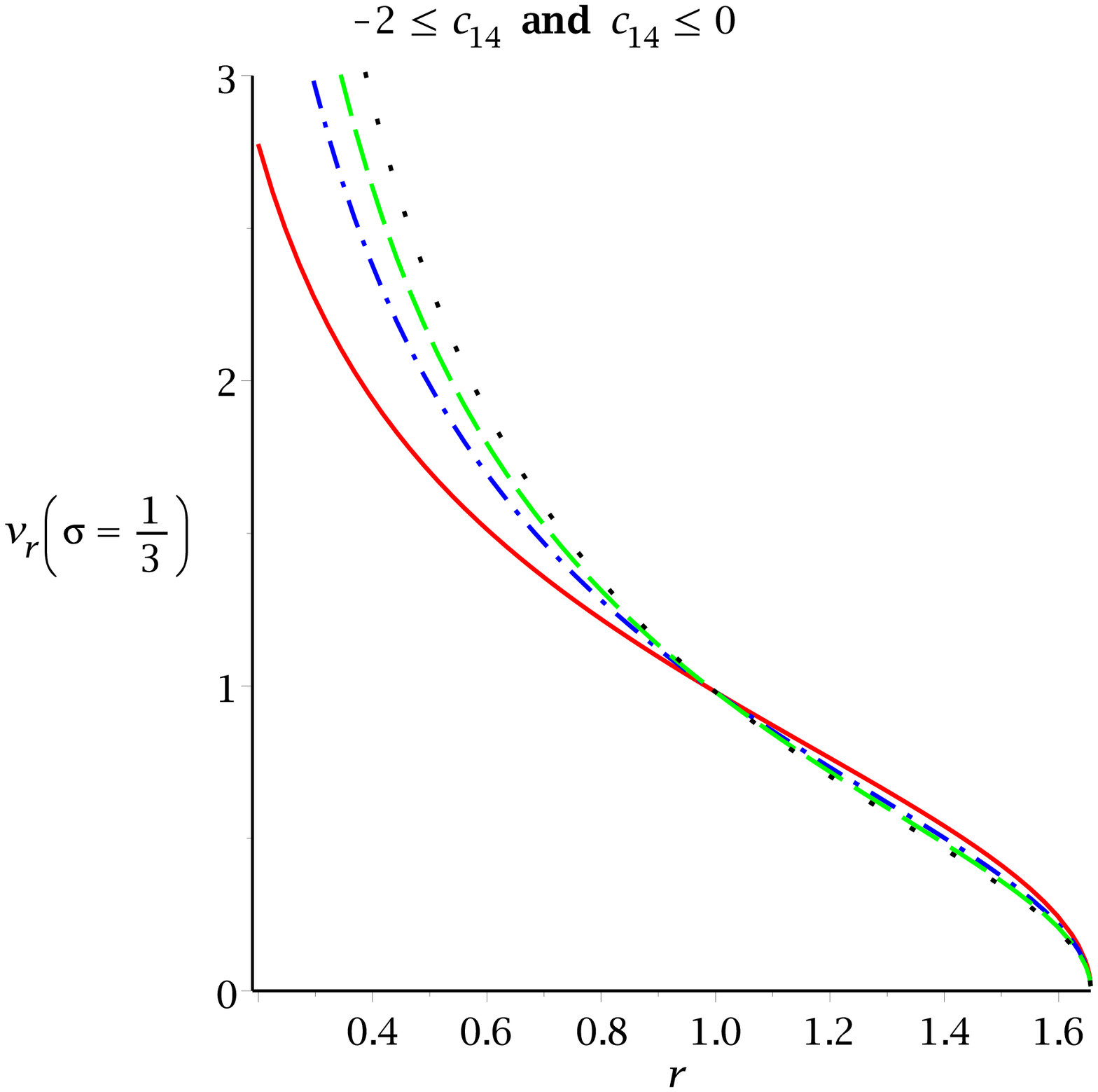}
\caption{These figures show $v_r$ for $\sigma=1/4$ and $1/3$.  
	For $\sigma=1/4$ or $\sigma=1/3$ and $0 \le c_{14} \le 2$ and $-2 \le c_{14} \le 0 $ we have:
	$c_{14}=\;0$ (red solid line), $c_{14}=\;\pm 1$ (blue dot-dashed line), 
	$c_{14}=\;\pm 3/2$ (green dashed line), $c_{14}=\;\pm 2$ (black dotted line).
	We assume $E=1.4$.}
	\label{vr025_033}	
\end{figure}
\begin{figure}[!ht]
\centering
	\includegraphics[width=7cm]{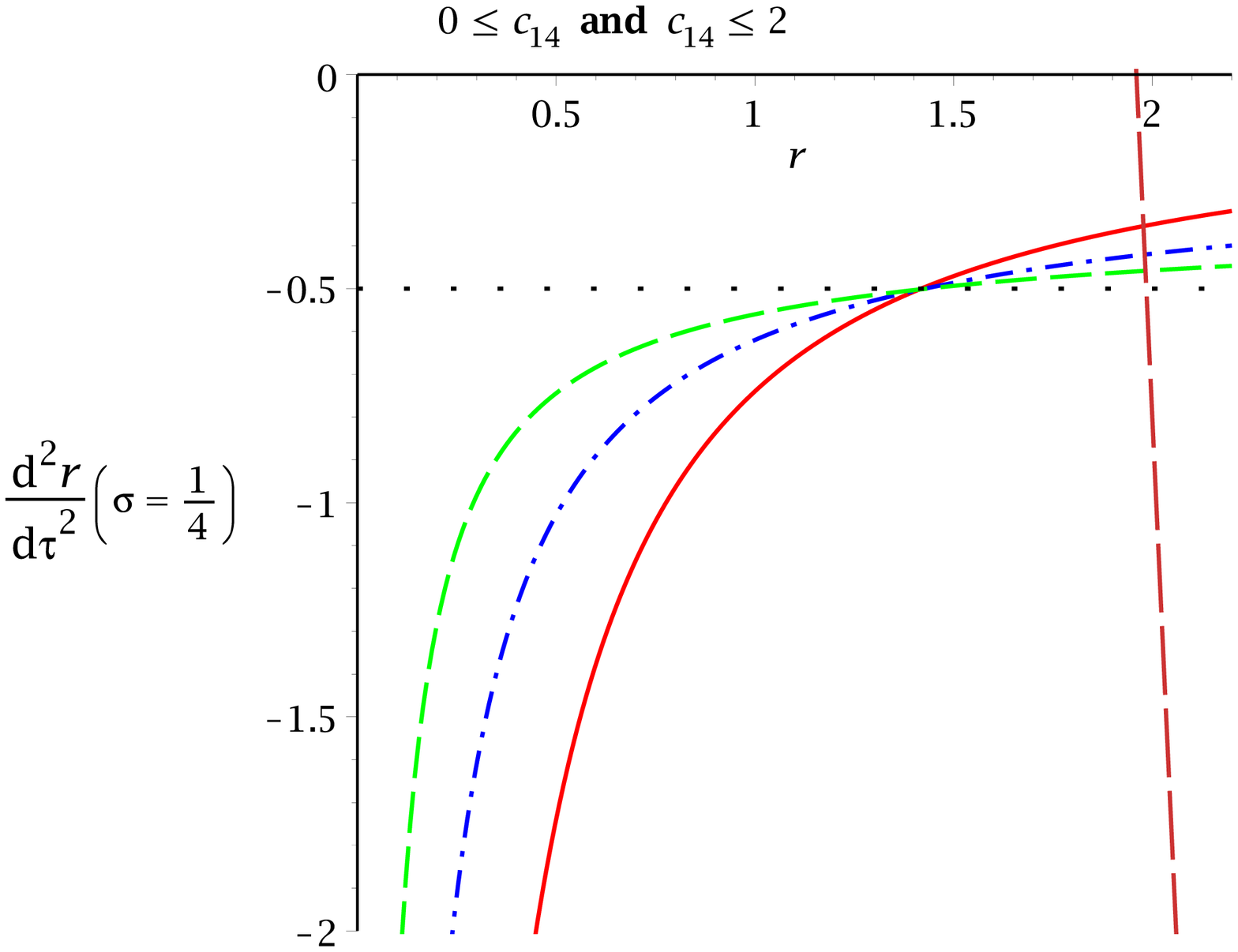}
	\includegraphics[width=7cm]{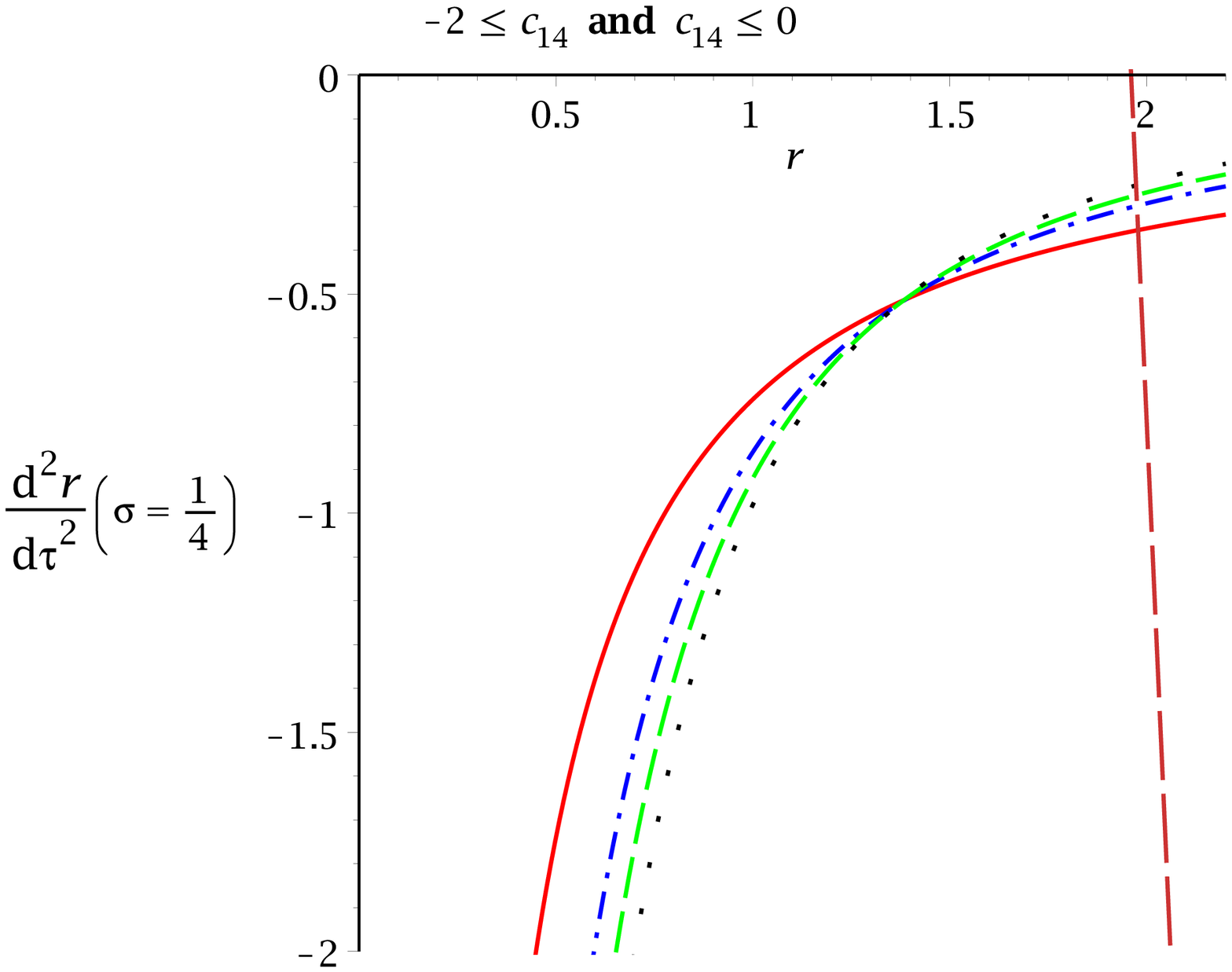}
	\includegraphics[width=7cm]{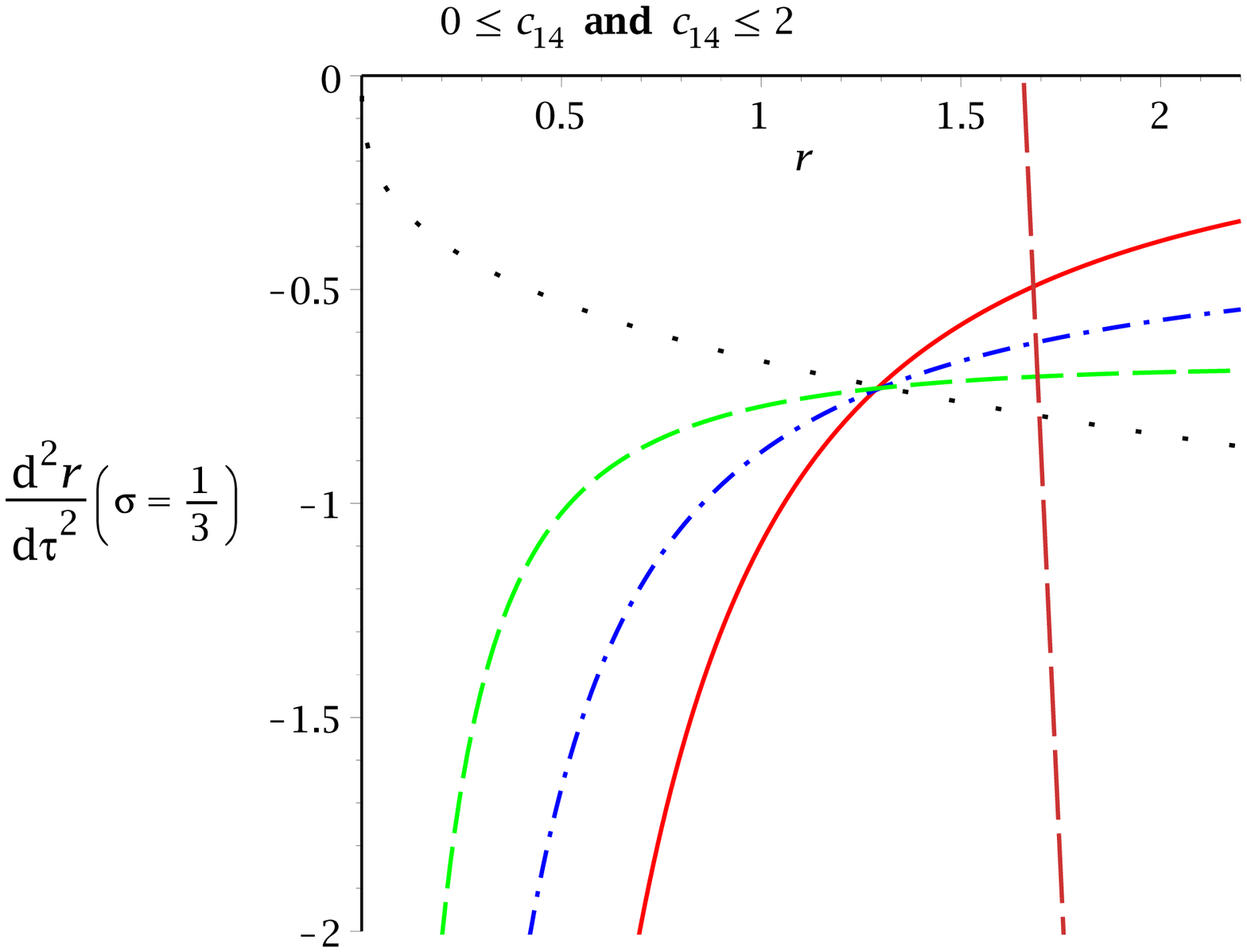}
	\includegraphics[width=7cm]{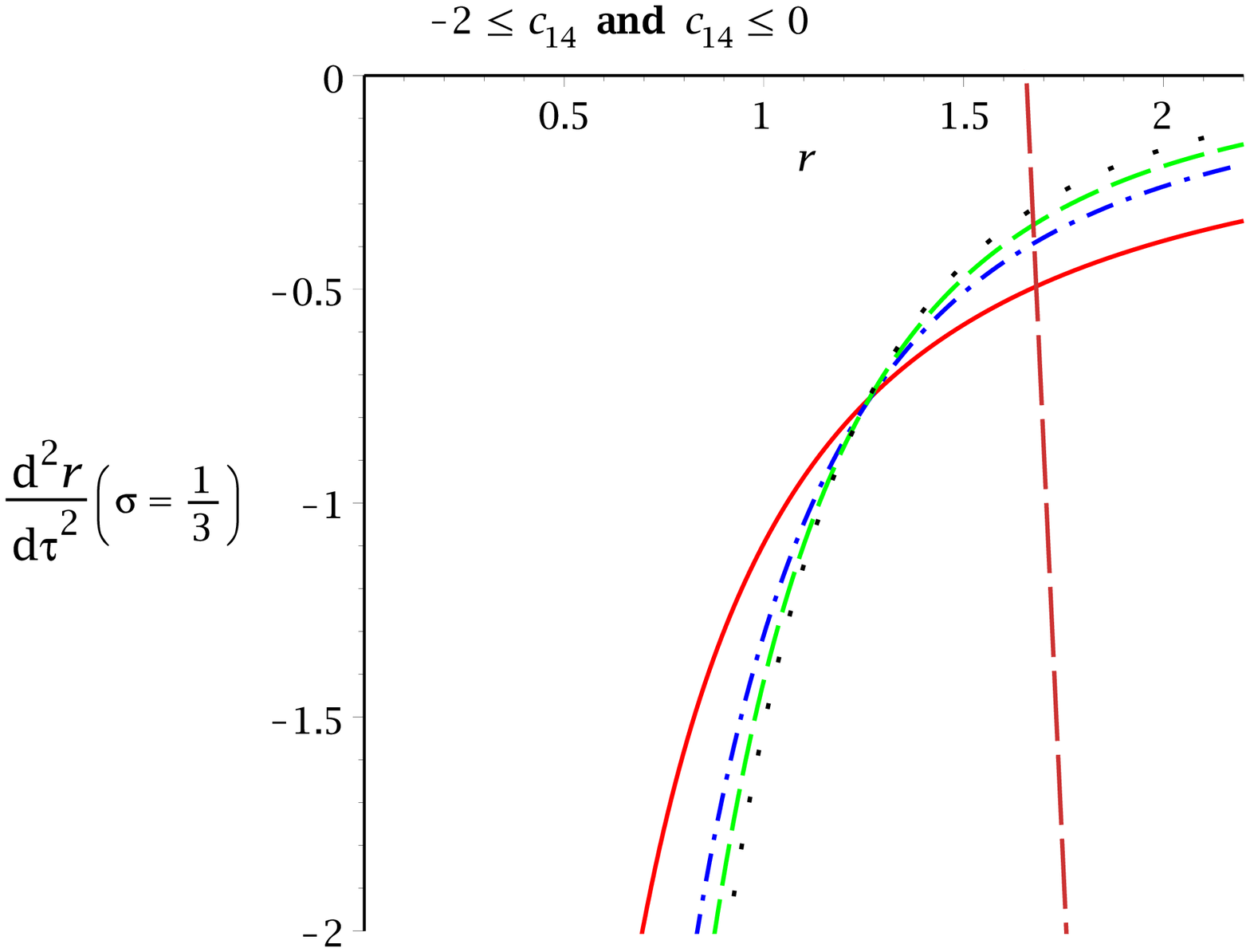}
\caption{These figures show $v_r$ for for $\sigma=1/4$ and $1/3$.  
	For $\sigma=1/4$ or $\sigma=1/3$ and $0 \le c_{14} \le 2$ and $-2 \le c_{14} \le 0$ 
we have:
	$c_{14}=\;0$ (red solid line), $c_{14}=\;\pm 1$ (blue dot-dashed line), 
	$c_{14}=\;\pm 3/2$ (green dashed line), $c_{14}=\;\pm 2$ (black dotted line).
	We assume $E=1.4$. The brown long-dashed line represent the $r_{max} \approx 1.96$
	for $\sigma=1/4$ and $r_{max} \approx 1.65$ for $\sigma=1/3$ }
	\label{d2r_025_033}	
\end{figure}
\begin{figure}[!ht]
\centering
	\includegraphics[width=7cm]{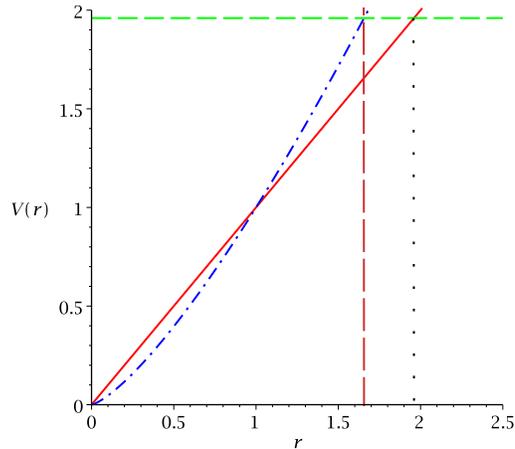}
\caption{This figure shows $V(r)$ assuming $P_z=L_z=0$ for for two values of 
	$\sigma=1/4$ (red solid line) 
	and $1/3$ (blue dot-dashed line).  
	We assume $E^2=1.96$ (green dashed line). The brown long-dashed line represent the $r_{max} \approx 1.96$
	for $\sigma=1/4$ and $r_{max} \approx 1.65$ for $\sigma=1/3$ (black dotted line).}
	\label{Vr_025_033}	
\end{figure}
	\begin{figure}[!ht]
\centering
	\includegraphics[width=7cm,angle=0]{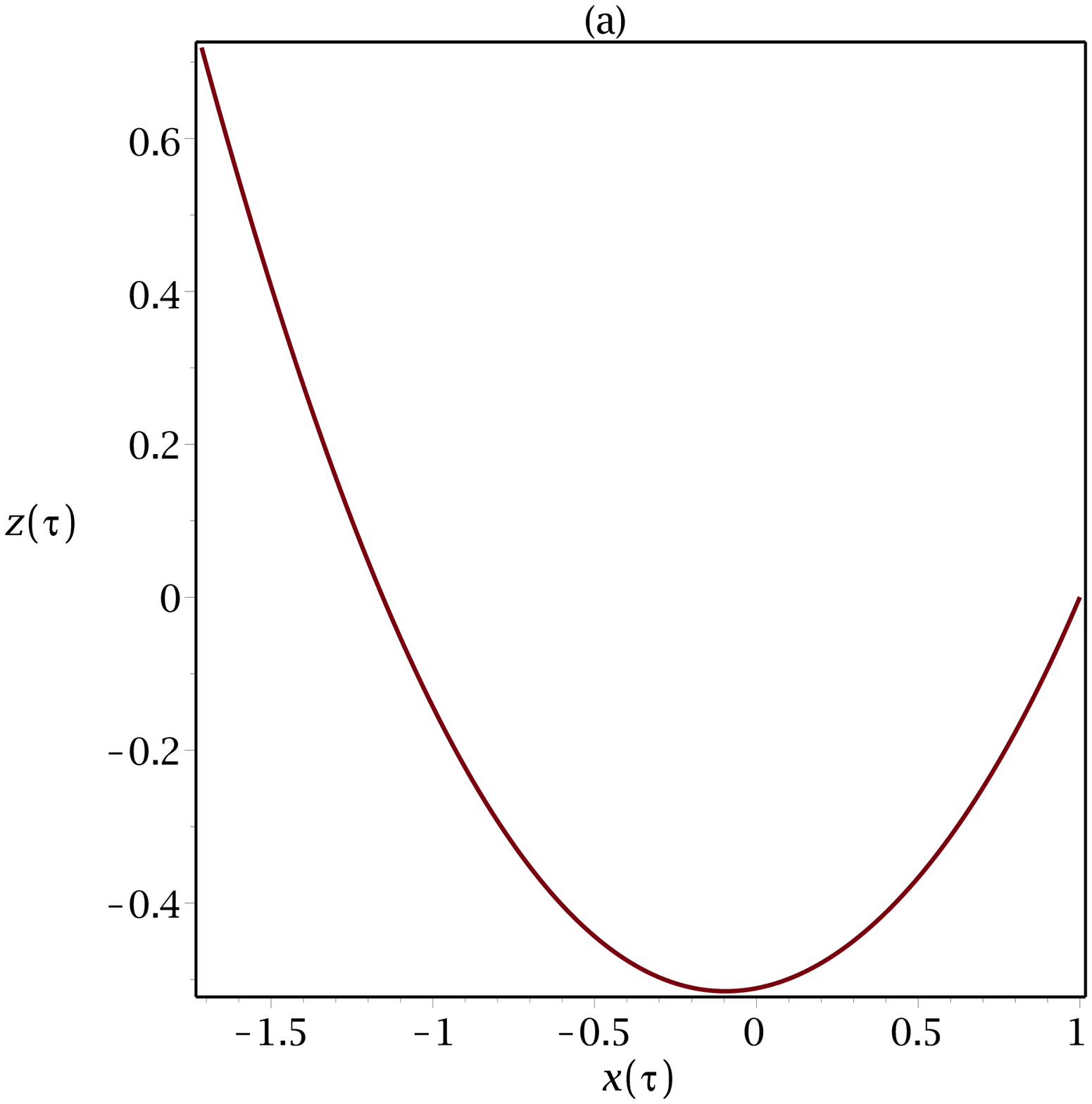}
	\includegraphics[width=7cm,angle=0]{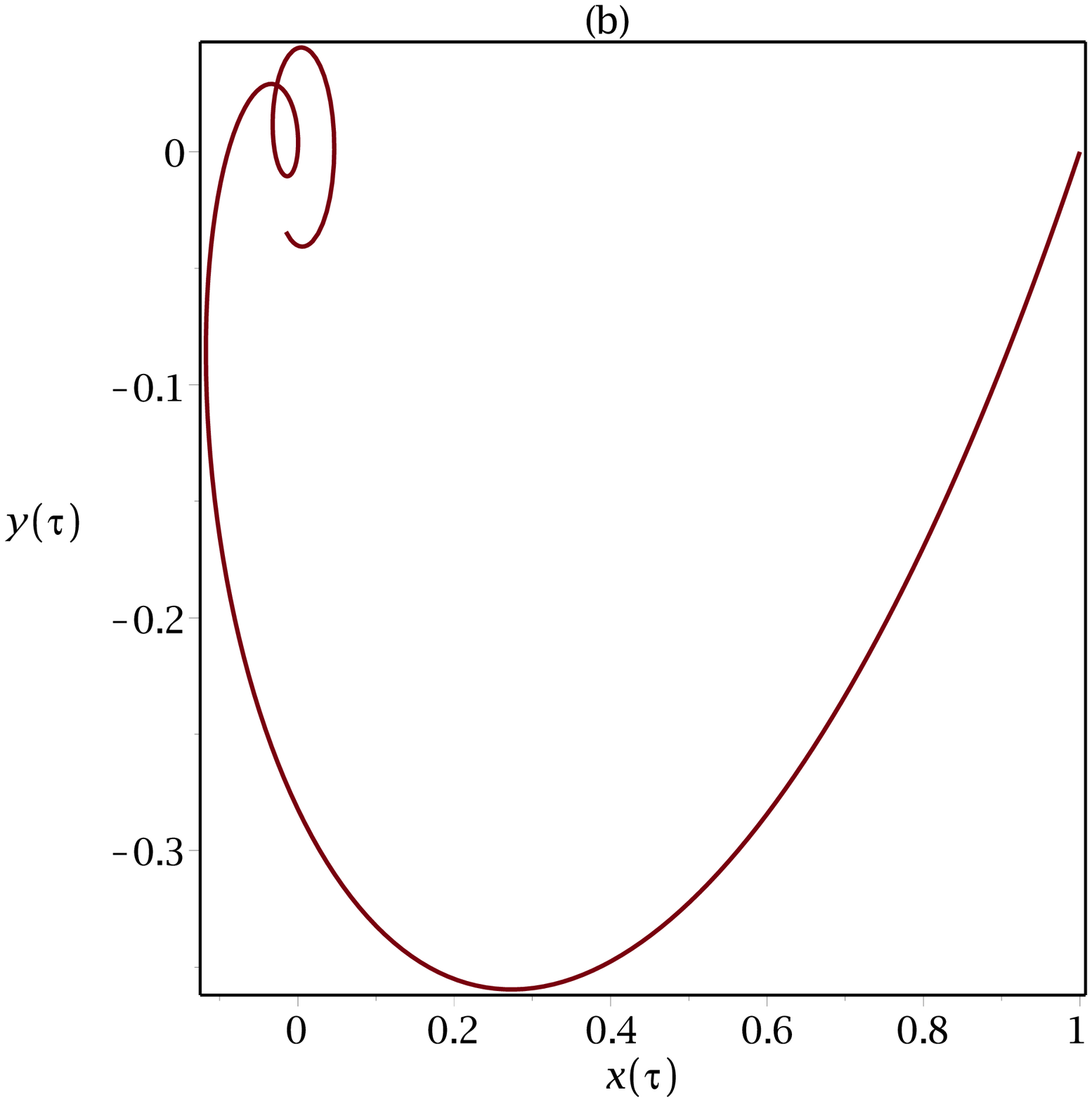}
\caption{The figure (a) shows the geodesic that depends on $r(\tau)$ and $z(\tau)$ in cylindrical
coordinates, projected in the plane $x(\tau) \times z(\tau)$ using that $\phi(\tau)=0$. 
We assume the initial conditions $r(0) = 1$, $z(0) = 0$, $t(0) = 0$, $\dot r(0) = -1$, 
$\dot z(0) = -1$, $\dot t(0) = 1$ for solving numerically the field equations.
The figure (b) shows the geodesic that depends on $r(\tau)$ and $\phi(\tau)$ in cylindrical
coordinates, projected in the plane $x(\tau) \times y(\tau)$ using that $z(\tau)=0$. 
We assume the initial conditions $r(0) = 1$, $\phi(0) = 0$, $t(0) = 0$, $\dot r(0) = -1$, 
$\dot \phi(0) = -1$, $\dot t(0) = 1$ for solving numerically the field equations.
We also assume that $a=1$, $c_{14}=1$, $\sigma=1/4$ and $E=1.4$.}
	\label{geod_rho_z_phi}	
\end{figure}	
	
Besides, no matter the value of $\sigma$, the radial velocity
for radii $ r <1$ and $0 \le c_{14} \le 2$ it is always smaller in EA 
than in GR ($c_{14}=0$), the bigger is $v_r$ the smaller is $c_{14}$.  
On the other hand, for radii $ r >1$ we have the opposite behavior, i.e., 
the radial velocity is always bigger in EA 
than in GR and the bigger is $v_r$ the bigger is $c_{14}$.
However, if we admit the interval $-2 \le c_{14} \le 0$, 
again, no matter the value of $\sigma$, we have  a similar pattern as before, but now the the relation between the GR and EA theories is inverted, although this 
range should not be considered in a healthy EA theory. 

From equation (\ref{vr}) it is easy to see that for $r=1$ the radial 
velocity is independent on $c_{14}$. 
This is the reason for the transition in the behavior of the radial 
velocity in the Figure \ref{vr025_033}.
If we restrict ourselves to the interval for which the EA theory is 
well behaved, that is, $0 \le c_{14} \le 2$, we can conclude that the 
vector field induces an increase in the speed of particles that move 
in the radial direction further away from the axis and a decrease 
in that speed for particles close to the axis of symmetry, 
both for particles that move away and for those that approach the axis, 
in comparison with the GR theory.

\section{Conclusions}

In this work we present, for the first time, as far as we know, all the possible solutions for a static cylindrical symmetric vacuum spacetime in the EA theory.
Besides the trivial flat solution, one of them is the generalization in EA theory of the LC spacetime in GR theory.

For this last one, we have noticed that, depending on the choice of $\sigma$ and $c_{14}$, the spacetime can be not singular at the axis $r=0$. More specifically, depending on the choice of $\sigma$ and $c_{14}$ in the range $\sigma>1/2$ and $c_{14}>3/2$, or $\sigma<1/2$ and $c_{14}>2$ the spacetime can be not singular at $r=0$. However, if $\sigma>1/2$ (this case the solution no longer preserves cylindrical symmetry) and if $c_{14}>2$ we are facing with a negative gravitational coupling constant, meaning a repulsive gravity.  Then, the result seems to be completely different from that one obtained with the GR theory where the axis $r=0$ is always singular for all values of $\sigma$ different from $0$, $1/2$ and $\infty$. Since the exponent $f$ is always negative if $c_{14}=0$ (GR limit), this is not true if we limit ourselves to the appropriate ranges for $\sigma$ and $c_{14}$. Meantime, this generalization has not the Rindler flat limit present in the LC solution in the GR theory, which reveals an important difference between the two theories.

We have also analyzed the geodesics properties of this generalized LC solution . The circular geodesics are the same of the GR theory, no matter the values of $c_{14}$. Although there is an  apparent  difference between the EA and GR theory, since in GR theory there are no trajectories parallel to the axis of symmetry for $0\leq\sigma\leq 1/2$, and it seems to be possible here, if we allows $c_{14}<0$. Nowadays, it is widely known that this implies in a bad behavior EA theory, because the presence of ghosts in the perturbed action. Thus, even geodesics in the $z$-directions are the same we have in the GR theory.

However, no matter the value of $\sigma$, the radial velocity for radii $ r <1$ and $0 \le c_{14} \le 2$ is always smaller in EA than in GR ($c_{14}=0$), the bigger is $v_r$ the smaller is $c_{14}$.  On the other hand, for radii $ r >1$ we have the opposite behavior, i.e., the radial velocity is always bigger in EA than in GR and the bigger is $v_r$ the bigger is $c_{14}$. Again, no matter the value of $\sigma$, the radial velocity for radii $r <1$ and $-2 \le c_{14} \le 0$ it is always bigger than in EA  than in GR ($c_{14}=0$). For radii $r >1$ we have the opposite behavior. Thus, in the range $0\leq\sigma\leq 1/2$ and  $0\leq c_{14}<2$, where we assure the cylindrical symmetry in a well healthy EA theory, the vector field seems to induce an increase in the speed of particles that move in the radial direction far away from the axis, in comparison with the GR theory. Besides, it decreases the particle speed close to the axis of symmetry, in comparison with the GR theory.

\section {Acknowledgments}

The financial assistance from FAPERJ/UERJ (MFAdaS) is gratefully acknowledged. The author (RC) acknowledges the financial support from FAPERJ (no.E-26/171.754/2000, E-26/171.533/2002 and E-26/170.951/2006). MFAdaS and RC also acknowledge the financial support from Conselho Nacional de Desenvolvimento Cient\'{\i}fico e Tecnol\'ogico - CNPq - Brazil.  The author (MFAdaS) also acknowledges the financial support from Financiadora de Estudos  Projetos - FINEP - Brazil (Ref. 2399/03). The authors are thankful to Dr. Anzhong Wang and Dr. Ted Jacobson for their comments and suggestions, which have led to an improved version of the article.

\appendix
\setcounter{section}{1}
\section*{Appendix A}

Using Maple 16 and GRTensorII we can also calculate the 18 scalar polynomial invariants of the Riemann tensor \cite{Carminati1991}\cite{Zakhary1997} in order to make sure of this result. They are: four real invariants from the Ricci tensor
(R, $r_1$, $r_2$, $r_2$); four complex invariants from the Weyl tensor ($w_1$, $w_2$), eight complex ($m_1$, $m_2$, $m_5$, $m_6$) and two real ($a^2L_z$, $m_4$) invariants from 
combinations of Ricci and Weyl, for this solution.
Thus, we have
\bqn
R &=& -4 \sigma^2 c_{14} r^{4 \sigma-8 \sigma^2+4 \sigma^2 c_{14}-2}=-4 \sigma^2 c_{14} r^{\frac{f}{2}},\nb\\
r_1 &=& 3 \sigma^4 c_{14}^2 r^f=3 \sigma^4 c_{14}^2 r^f, \nb\\
r_2 &=& 3 \sigma^6 c_{14}^3 r^{12 \sigma-24 \sigma^2+12 \sigma^2 c_{14}-6}=3 \sigma^6 c_{14}^3 r^{\frac{3f}{2}},\nb\\
r_2 &=& \frac{21}{4} \sigma^8 c_{14}^4 r^{16 \sigma-32 \sigma^2+16 \sigma^2 c_{14}-8}=\frac{21}{4} \sigma^8 c_{14}^4 r^{2f},\nb\\
\Re(w_1) &=& \frac{8}{3} \sigma^2 r^{8 \sigma-16 \sigma^2+8 \sigma^2 c_{14}-4} \times \nb\\
&&\left[ (\sigma^2-6 \sigma^3+12 \sigma^4)  c_{14}^2+(3 \sigma-48 \sigma^4+48 \sigma^3-18 \sigma^2) c_{14}+ \right.\nb\\
&&\left. (3-18 \sigma+48 \sigma^4-72 \sigma^3+48 \sigma^2) \right]\nb\\
&=&\frac{8}{3} \sigma^2 r^f \left[ (\sigma^2-6 \sigma^3+12 \sigma^4)  c_{14}^2+(3 \sigma-48 \sigma^4+48 \sigma^3-18 \sigma^2) c_{14}+ \right.\nb\\
&&\left. (3-18 \sigma+48 \sigma^4-72 \sigma^3+48 \sigma^2) \right],\nb\\
\Im(w_1) &=& 0,\nb\\
\Re(w_2) &=& -\frac{8}{9} \sigma^4 r^{12 \sigma-24 \sigma^2+12 \sigma^2 c_{14}-6} \times \nb\\
&&\left[ (36 \sigma^4-18 \sigma^3+2 \sigma^2)  c_{14}^3-
(-360 \sigma^4+9 \sigma-108 \sigma^2+360 \sigma^3) 
 c_{14}^2-\right.\nb\\
&&\left. (792 \sigma^2+9-1512 \sigma^3-162 \sigma+1008 \sigma^4) 
 c_{14}- \right.\nb\\
&&\left. (-54-1296 \sigma^2-864 \sigma^4+1728 \sigma^3+432 \sigma) \right]\nb\\
&=&-\frac{8}{9} \sigma^4 r^{\frac{3f}{2}} \times \nb\\
&&\left[ (36 \sigma^4-18 \sigma^3+2 \sigma^2)  c_{14}^3-
(-360 \sigma^4+9 \sigma-108 \sigma^2+360 \sigma^3) c_{14}^2-\right.\nb\\
&&\left. (792 \sigma^2+9-1512 \sigma^3-162 \sigma+1008 \sigma^4) c_{14}- \right.\nb\\
&&\left. (-54-1296 \sigma^2-864 \sigma^4+1728 \sigma^3+432 \sigma) \right],\nb\\
\Im(w_2) &=& 0,\nb\\
\Re(m_1) &=& 0,\nb\\
\Im(m_1) &=& 0,\nb\\
\Re(m_2) &=& \frac{8}{3} \sigma^6 r^{16 \sigma-32 \sigma^2+16 \sigma^2 c_{14}-8}\times\nb\\
&& \left[ (\sigma^2-6 \sigma^3+12 \sigma^4)  c_{14}^4+ (3 \sigma-48 \sigma^4+48 \sigma^3-18 \sigma^2)  c_{14}^3+ \right.\nb\\
&&\left. (3-18 \sigma+48 \sigma^4-72 \sigma^3+48 \sigma^2) c_{14}^2 \right]\nb\\
&=&\frac{8}{3} \sigma^6 r^{2f}\times\nb\\
&& \left[ (\sigma^2-6 \sigma^3+12 \sigma^4)  c_{14}^4+ (3 \sigma-48 \sigma^4+48 \sigma^3-18 \sigma^2)  c_{14}^3+ \right.\nb\\
&&\left. (3-18 \sigma+48 \sigma^4-72 \sigma^3+48 \sigma^2) c_{14}^2 \right],\nb\\
\Im(m_2) &=& 0,\nb\\
m_3 &=& \frac{8}{3} \sigma^6 r^{16 \sigma-32 \sigma^2+16 \sigma^2 c_{14}-8} \times\nb\\
&&\left[(\sigma^2-6 \sigma^3+12 \sigma^4)  c_{14}^4+(3 \sigma-48 \sigma^4+48 \sigma^3-18 \sigma^2)  c_{14}^3+ \right.\nb\\
&&\left. (3-18 \sigma+48 \sigma^4-72 \sigma^3+48 \sigma^2)  c_{14}^2\right]\nb\\
&=&\frac{8}{3} \sigma^6 r^{2f} \times\nb\\
&&\left[(\sigma^2-6 \sigma^3+12 \sigma^4)  c_{14}^4+(3 \sigma-48 \sigma^4+48 \sigma^3-18 \sigma^2)  c_{14}^3+ \right.\nb\\
&&\left. (3-18 \sigma+48 \sigma^4-72 \sigma^3+48 \sigma^2)  c_{14}^2\right],\nb\\
m_4 &=& -\frac{4}{3} \sigma^8 r^{-10+20 \sigma-40 \sigma^2+20 \sigma^2 c_{14}} \times\nb\\
&&\left[ (\sigma^2-6 \sigma^3+12 \sigma^4)  c_{14}^5- (3 \sigma-48 \sigma^4+48 \sigma^3-18 \sigma^2)  c_{14}^4-\right. \nb\\
&&\left. (3-18 \sigma+48 \sigma^4-72 \sigma^3+48 \sigma^2) c_{14}^3\right]\nb\\
&=&-\frac{4}{3} \sigma^8 r^{\frac{5f}{2}} \times\nb\\
&&\left[ (\sigma^2-6 \sigma^3+12 \sigma^4)  c_{14}^5- (3 \sigma-48 \sigma^4+48 \sigma^3-18 \sigma^2)  c_{14}^4-\right. \nb\\
&&\left. (3-18 \sigma+48 \sigma^4-72 \sigma^3+48 \sigma^2) c_{14}^3\right],\nb\\
\Re(m_5) &=& -\frac{8}{9} \sigma^8 r^{-10+20 \sigma-40 \sigma^2+20 \sigma^2 c_{14}} \times \nb\\
&&\left[ 2 \sigma^2 (3 \sigma-1) (6 \sigma-1) c_{14}^5+9 \sigma (2 \sigma-1) (20 \sigma^2-10 \sigma+1) c_{14}^4+\right.\nb\\
&&\left. 9 (28 \sigma^2-14 \sigma+1) (2 \sigma-1)^2 c_{14}^3+54 (2 \sigma-1)^4 c_{14}^2\right]\nb\\
&=&\frac{8}{9} \sigma^8 r^{\frac{5f}{2}} \times \nb\\
&&\left[ 2 \sigma^2 (3 \sigma-1) (6 \sigma-1) c_{14}^5-9 \sigma (2 \sigma-1) (20 \sigma^2-10 \sigma+1) c_{14}^4+\right.\nb\\
&&\left. 9 (28 \sigma^2-14 \sigma+1) (2 \sigma-1)^2 c_{14}^3-54 (2 \sigma-1)^4 c_{14}^2\right],\nb\\
\Im(m_5) &=& 0,\nb\\
\Re(m_6) &=& 0,\nb\\
\Im(m_6) &=& 0,\nb\\
\eqn
where the symbol $\Re$ and $\Im$ denote the real and imaginary parts of the scalar.

\section{References}

\end{document}